\documentclass[aps,pra,onecolumn,groupedaddress]{revtex4-1}
\usepackage{graphicx}
\usepackage{setspace}
\usepackage{epsfig}
\usepackage{amsmath}
\usepackage{graphicx}
\usepackage{textcomp}
\usepackage{amssymb}
\usepackage{slashbox}
\usepackage{geometry}
\usepackage{subfigure}
\usepackage{ulem}
\usepackage{color}
\usepackage{float}
\usepackage{comment}

\newcommand{\nn}{\nonumber}
\newcommand{\GX}{\Gamma_{X}(\mathbf r)}
\newcommand{\GXX}{\Gamma_{XX}(\mathbf r)}
\newcommand{\Gi}{\Gamma_{i}(\mathbf r)}
\newcommand{\pd}{\partial}
\newcommand{\barGX}{\bar{\Gamma}_{X}}
\newcommand{\barGXX}{\bar{\Gamma}_{XX}}
\newcommand{\rPOS}{\mathbf r}

\begin{document}
\title{Overcoming Auger recombination in nanocrystal quantum dot laser using spontaneous emission enhancement}
\author{Shilpi Gupta and Edo Waks}
\affiliation{Department of Electrical and Computer Engineering,\\
Institute for Research in Electronics and Applied Physics, and\\ 
Joint Quantum Institute,\\
University of Maryland College Park,\\
Maryland 20742, USA}

\begin{abstract}
We propose a method to overcome Auger recombination in nanocrystal quantum
dot lasers using cavity-enhanced spontaneous emission. We derive a
numerical model for a laser composed of nanocrystal quantum dots coupled to
optical nanocavities with small mode-volume. Using this model, we
demonstrate that spontaneous emission enhancement of the biexciton
transition lowers the lasing threshold by reducing the effect of Auger
recombination. We analyze a photonic crystal nanobeam cavity laser as a
realistic device structure that implements the proposed approach.
\end{abstract}

\maketitle

\section{Introduction}
\label{section:Introduction}

Room-temperature nanolasers have applications in fields ranging from optical
communications and information processing~\cite{Hill2009} to biological
sensing ~\cite{Kita2011} and medical diagnostics~\cite{Gourley2005}.
Colloidally synthesized nanocrystal quantum dots are a promising gain
material for nanolasers. These quantum dots are efficient emitters at room
temperature \cite{Alivisatos2003,Qu2002}, have broadly tunable emission
frequencies~\cite{Alivisatos1996,KlimovBook2003} and are easy to integrate
with photonic structures~\cite{Eisler2002,Snee2005,Min2006}.

Nanocrystal quantum dot lasers have been demonstrated using resonant
structures such as distributed feedback gratings~\cite{Eisler2002},
microspheres~\cite{Snee2005}, and micro-toroids~\cite{Min2006}. However,
these devices have exhibited high lasing thresholds due to fast non-radiative
decay caused by Auger recombination~\cite{Klimov2000,Klimov2000a}.
Nanocrystal quantum dots have a fast Auger recombination rate owing to the
tight spatial confinement of carriers~\cite{Klimov2000a}. One approach to
reduce Auger recombination is by engineering quantum dots with decreased
spatial confinement. For example, elongated nanocrystals (quantum rods) can
reduce Auger recombination \cite{HtoonPRL2003,HtoonAPL2003} to achieve lower
threshold lasing ~\cite{Kazes2002}. Core/shell heteronanocrystals may also
reduce the carrier spatial confinement~\cite{Ivanov2004,Nanda2006}, but have
yet to be successfully integrated into a laser structure.

Here we show that spontaneous emission rate enhancement in a small mode
volume cavity ~\cite{Purcell1946} can overcome Auger recombination and enable low threshold lasing. We derive a model for a nanocrystal quantum dot
laser using a master equation formalism that accounts for both Auger
recombination and spontaneous emission enhancement. Using this model we show
that spontaneous emission enhancement reduces the effect of Auger
recombination, resulting in up to a factor of 17 reduction in the lasing
threshold. We analyze a nanobeam photonic crystal cavity as a promising device implementation to achieve low threshold lasing in the presence of Auger recombination.

In section \ref{section:NumericalModel} we derive the theoretical formalism
for a nanocrystal quantum dot laser. Section \ref{section:UFA} presents
numerical calculations for a general cavity structure under the uniform-field
approximation. In section \ref{section:NanoBeamCavity} we propose and analyze
a nanobeam photonic crystal cavity design as a potential device
implementation of a nanocrystal quantum dot laser.

\section{Derivation of numerical model}
\label{section:NumericalModel}
\begin{figure}[htbp]
\centering
\includegraphics[width=10cm]{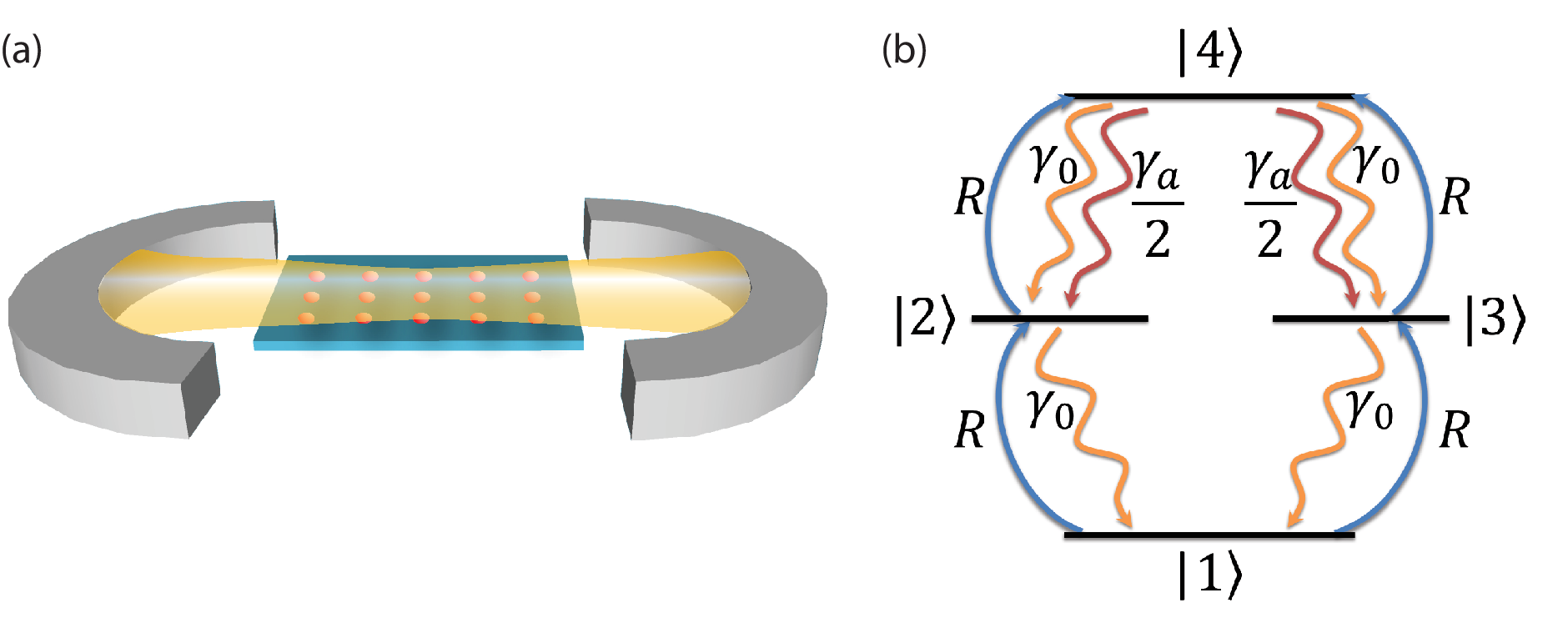}
\caption{(a) Schematic of a laser composed of nanocrystal quantum dots coupled to an optical cavity. (b) Level diagram for a four-level model of a nanocrystal quantum dot.}
\label{fig:Fig1_cavity_NQD}
\end{figure}

Figure~\ref{fig:Fig1_cavity_NQD}(a) illustrates the general model for a nanocrystal
quantum dot laser. The laser is composed of an ensemble of quantum dots
coupled to a single cavity mode. The level structure of the quantum dots,
shown in Fig.~\ref{fig:Fig1_cavity_NQD}(b), consists of four states: a ground state
$|1\rangle$ which contains no carriers, the single exciton states $|2\rangle$
and $|3\rangle$ which contain a single electron-hole pair, and the biexciton
state $|4\rangle$ which contains two electron-hole pairs. In the single
exciton states, the quantum dot absorbs and emits a photon with nearly equal
probability.  Thus, only the biexciton state can provide optical gain
~\cite{Klimov2003}. However, this state suffers from Auger recombination
where an electron-hole pair recombines and transfers energy non-radiatively
to a third carrier~\cite{Klimov2000a}. The strong carrier confinement in the
quantum dots leads to fast Auger recombination, resulting in a low biexciton
radiative efficiency.

Figure ~\ref{fig:Fig1_cavity_NQD}(b) also shows the relevant decay rates for our
quantum dot model. The biexciton state decays to each exciton state with the
rate $\gamma_2 = \gamma_0 + \gamma_a/2$, where $\gamma_0$ is the spontaneous
emission rate and $\gamma_a$ is the total Auger recombination rate of the
biexciton state. We assume the single exciton states decay predominantly by
spontaneous emission.  We also assume equal spontaneous emission rates for
all four allowed transitions, and ignore long-lived trap states that are
responsible for blinking behavior~\cite{Nirmal1996,Jones2003}. These states
can be incorporated as additional energy levels in the model.  The quantum
dot is incoherently pumped with an external source characterized by the
excitation rate $R$.

In bare nanocrystal quantum dots Auger recombination is an order of magnitude
faster than spontaneous emission~\cite{Klimov2000a}. It therefore dominates
the decay of the biexciton state and quenches the optical gain. However, when
the quantum dot spectrally couples to an optical cavity, its spontaneous
emission rate increases by the factor ~\cite{Gerard2003}.
\begin{equation}\label{eq:PurcellFactor}
F(\rPOS_0) =  1 + \frac{2g^2(\rPOS_0)}{\gamma_0K_{XX}}
\end{equation}
where $g(\rPOS_0)$ is the cavity-quantum dot coupling strength given by
\begin{equation}\label{eq:CouplingStrength}
g(\rPOS_0)= \frac{\mathbf{\mu}\cdot\mathbf{\hat{e}}}{\hbar}\sqrt{\frac{\hbar\omega_c}{2 \epsilon_{0}V_m}}\frac{|E(\rPOS_0)|}{|E(\rPOS)|_{max}}
\end{equation}
Here, $E(\rPOS_0)$ is the electric field amplitude, $\mathbf{\hat{e}}$ is
the polarization direction of the cavity mode at the quantum dot position
$\rPOS_0$, $\omega_c$ is the cavity mode resonant frequency, $V_m = \int d^3\rPOS
\epsilon(\rPOS) |E(\rPOS)|^2/[|E(\rPOS)|^2]_{max}$ is the cavity mode-volume
~\cite{Zhang2011}, $\epsilon_{0}$ is the permittivity of free space,
$\epsilon(\rPOS)$ is the relative dielectric permittivity and $\mathbf{\mu}$
is the quantum dot dipole moment. The rate $K_{XX}= (\gamma_0 + 2\gamma_2 + \gamma_{d})/2$ represents the total linewidth of the biexciton state, which is dominated by the dephasing rate $\gamma_{d}$ at room-temperature
~\cite{VanSark2001,Lounis2000,Klimov2000a}. We note that
Eq.~(\ref{eq:PurcellFactor}) is different from the more common expression for
$F$ that depends on the ratio of the cavity quality factor $Q$ and the cavity
mode-volume $V_m$~\cite{Purcell1946,Gerard2003}. This difference occurs
because at room temperature the dephasing rate of nanocrystal quantum dots is
much larger than the cavity linewidth. The device therefore operates in the
bad emitter regime, where $F$ becomes independent of the cavity $Q$. By
engineering cavities with small mode-volumes, we can achieve large $F$ and
enhance the spontaneous emission rate, thereby increasing the radiative
efficiency of the quantum dot in the presence of Auger recombination.

To analyze the nanocrystal quantum dot laser in the presence of Auger
recombination and spontaneous emission enhancement, we begin with the master
equation
\begin{equation}\label{eq:MasterEquation}
\frac{\partial\rho}{\partial t} = \frac{i}{\hbar}[\rho, \mathbf{H}] + \mathbf{L}\rho
\end{equation}
where $\rho$ is the density matrix of the combined cavity-quantum dot system,
$\mathbf{H}$ is the Hamiltonian, and $\mathbf{L}$ is the Liouvillian
superoperator that accounts for incoherent damping and excitation processes.
The Hamiltonian of the system is given by $\mathbf{H_{cavity}} + \mathbf{H_{NQD}}
+ \mathbf{H_{JC}}$, where
\begin{eqnarray}
\label{eq:H_cavity}
\mathbf{H_{cavity}} &=& \hbar\omega_{c}\mathbf{a}^\dagger\mathbf{a}\\
\label{eq:H_NQD}
\mathbf{H_{NQD}} &=& \sum_{m=1}^{N}\hbar\omega^{X}_{m}(\sigma_{22,m} + \sigma_{33,m}) + \hbar\omega^{XX}_{m}\sigma_{44,m}\\
\label{eq:H_JC}
\mathbf{H_{JC}} &=& \sum_{i=m}^{N} \hbar g^{X}_{m}(\rPOS_m)(\sigma_{21,m}\mathbf{a} + \sigma_{12,m} \mathbf{a}^\dagger + \sigma_{31,m}\mathbf{a} + \sigma_{13,m} \mathbf{a}^\dagger)\nn\\
&+& \hbar g^{XX}_{m}(\rPOS_m)(\sigma_{42,m}\mathbf{a} + \sigma_{24,m} \mathbf{a}^\dagger + \sigma_{43,m}\mathbf{a} + \sigma_{34,m} \mathbf{a}^\dagger)
\end{eqnarray}
In the above equations $\mathbf{a}$ and $\mathbf{a}^\dagger$ are the bosonic
annihilation and creation operators of the cavity mode. The summation is
carried out over all quantum dots in the cavity, where we denote the total
number of quantum dots by $N$.  For the $m^{th}$ quantum dot, $\sigma_{jk,m}
= |j\rangle\langle k|$ represents the atomic dipole operator when $j \neq k$
and the atomic population operator when $j = k$, for the single exciton
states ($j = 2, 3$) and the biexciton state ($j = 4$). We set the energy of
the quantum dot ground state to zero. We define $\omega^{X}_{m}$ and
$\omega^{XX}_{m}$ as the resonant frequencies of the single-exciton and
biexciton transitions, respectively.  Similarly, the cavity-quantum dot
coupling strengths for the exciton and biexciton transitions are $g^{X}_{m}(\rPOS_m)$
and $g^{XX}_{m}(\rPOS_m)$ for the $m^{th}$ quantum dot at position $\rPOS_m$. At room
temperature, the homogenous linewidth of these quantum dots is much larger
than the biexcitonic shift \cite{Caruge2004,Empedocles1996,Empedocles1999,Norris1996}. We therefore assume all four
transitions of each quantum dot are resonantly coupled to the cavity mode
($\omega_{c} = \omega^{X}_{m} = \omega^{XX}_{m}/2$). The Liouvillian
superoperator $\mathbf{L}$ is fully defined in Appendix
~\ref{appendix:Liouvillian}.

The master equation is difficult to solve both analytically and numerically
when the number of quantum dots becomes large.  However, we can simplify the
calculations by applying the semi-classical approximation in which the
coherence between the atoms and the field is neglected
~\cite{WallsMilburn2007,Benson1999} and the density matrix can be factorized into a
product of the state of the field and atoms (see Appendix
~\ref{appendix:Equations_N}). Under this approximation, the system is
described by the average cavity photon number, $p$, and the quantum dot
population density, $n_j(\rPOS) = \lim_{\Delta V\to
0}\sum_m\langle\sigma_{jj}^m\rangle/\Delta V$, where the sum is carried out
over all quantum dots contained in a small volume $\Delta V$ at location
$\rPOS$. We note that $n_j(\rPOS)$ is a function of the position $\rPOS$ inside the
cavity because of the non-uniform cavity field distribution. We derive the
equations of motion of $n_j(\rPOS)$ from the master equation (see Appendix
~\ref{appendix:Equations_N}) as
\begin{eqnarray}
\label{eq:ThreeLevel_N1}
{\pd n_1(\rPOS)\over\pd t} &=& \GX[(p + 1)(n_2(\rPOS) + n_3(\rPOS)) - 2pn_1(\rPOS)] + \gamma_0 [n_2(\rPOS) + n_3(\rPOS)] - 2Rn_1(\rPOS)\\
\label{eq:ThreeLevel_N2}
{\pd n_2(\rPOS)\over\pd t}  &=& - \GX[(p + 1)n_2(\rPOS) - pn_1(\rPOS)] + \GXX[(p + 1)n_4(\rPOS) - pn_2(\rPOS)]\nn\\ &-& \gamma_0 n_2(\rPOS) + \gamma_2 n_4(\rPOS) + R[n_1(\rPOS) - n_2(\rPOS)]\\
\label{eq:ThreeLevel_N3}
{\pd n_3(\rPOS)\over\pd t}  &=& - \GX[(p + 1)n_3(\rPOS) - pn_1(\rPOS)] + \GXX [(p + 1)n_4(\rPOS) - pn_3(\rPOS)]\nn\\ &-& \gamma_0 n_3(\rPOS) + \gamma_2 n_4(\rPOS) + R[n_1(\rPOS) - n_3(\rPOS)]\\
\label{eq:ThreeLevel_N4}
{\pd n_4(\rPOS)\over\pd t} &=& - \GXX[2(p + 1)n_4(\rPOS) - p(n_2(\rPOS)+n_3(\rPOS))] - 2\gamma_2 n_4(\rPOS) \nn\\ &+& R[n_2(\rPOS)+n_3(\rPOS)]
\end{eqnarray}

In the above equations, $\GX = 2g^2(\rPOS)/K_X$ and $\GXX = 2g^2(\rPOS)/K_{XX}$ are
the modified spontaneous emission rates of the single-exciton and biexciton
transitions,  where $K_{X} = (\gamma_0 + \gamma_{d} + 3R)/2$ and $K_{XX} =
(\gamma_0 + 2\gamma_2 + \gamma_{d} + R)/2$. Here, we assume equal coupling
strength for the single-exciton and biexciton transitions. We also treat the
quantum dots in a small volume $\Delta V$ of the cavity to be identical, and
therefore drop the subscript $m$ from the coupling strength ($g(\rPOS)$ =
$g^{X}_{m}(\rPOS_m)$ = $g^{XX}_{m}(\rPOS_m)$). 

The average cavity photon number satisfies a rate equation given by (see
Appendix ~\ref{appendix:Equations_n} for derivation)
\begin{equation}\label{eq:ThreeLevel_n}
{\pd p \over\pd t}  = -p\kappa  + pG(p) + \alpha(p)
\end{equation}
where $\kappa = \omega_c/Q$ is the cavity energy decay rate.  The above
equation is coupled to the quantum dot population density rate equations
through the cavity gain coefficient
\begin{equation}\label{eq:G}
G(p) = \int d^3\rPOS\left\{\GX[n_2(\rPOS) + n_3(\rPOS) - 2n_1(\rPOS)] + \GXX[2n_4(\rPOS) - n_2(\rPOS) - n_3(\rPOS)]\right\}
\end{equation}
and the spontaneous emission rate into the lasing mode
\begin{equation}\label{eq:alpha}
\alpha(p) = \int d^3\rPOS\left\{\GX[n_2(\rPOS) + n_3(\rPOS)] + 2\GXX n_4(\rPOS)\right\}
\end{equation}
where the integral is over all space. We use the
notation $G(p)$ and $\alpha(p)$ to highlight the fact that the above
coefficients have a $p$ dependence because the atomic densities $n_j(\rPOS)$
depend on the cavity photon number. The absorbed pump power of the
nanocrystal quantum dot laser is given by
\begin{equation}
\label{eq:P_abs}
P_{abs} = \hbar\omega_{p}R\int_{V_p} d^3\rPOS[2n_1(\rPOS) + n_2(\rPOS) + n_3(\rPOS)]
\end{equation}
where $\omega_{p}$ is the pump frequency and $V_p$ is the optically pumped
volume. The output power of the laser is given by
\begin{equation}
\label{eq:P_out}
P_{out} = \hbar\omega_{c}p\kappa
\end{equation}

An important figure of merit for small mode-volume cavities is the
spontaneous emission coupling efficiency, denoted by $\beta$. This parameter
quantifies the fraction of photons spontaneously emitted to the cavity mode.
A $\beta$ approaching unity achieves thresholdless lasing ~\cite{Bjork1991}.
In the quantum dot model,  the single exciton and biexciton transitions have
different coupling efficiencies given by
\begin{eqnarray}
\label{eq:Beta_X}
\beta_X(\rPOS) &=&\frac{\GX}{\GX + \gamma_0}\\
\label{eq:Beta_XX}
\beta_{XX}(\rPOS)  &=&\frac{\GXX}{\GXX + \gamma_2}
\end{eqnarray}
The above coupling efficiencies depend on the position $\rPOS$ due to the
spatially varying cavity field intensity.  The rate equations Eqs.
(\ref{eq:ThreeLevel_N1})-(\ref{eq:ThreeLevel_n}) describe the dynamics of a
general nanocrystal quantum dot laser. We will use these equations in the
remaining sections.

\section{Lasing analysis under uniform-field approximation}
\label{section:UFA}

The general cavity-quantum dot rate equation model, developed in the previous
section, is still difficult to solve due to the spatial variation of the
coupling strength $g(\rPOS)$. This spatial variation leads to a complex set of
coupled differential equations for each position inside the cavity
volume. We note that this complexity is not unique to the system we study. It
occurs in virtually all laser systems and is responsible for effects such as
spatial hole burning ~\cite{MilonniEberlyBook}. One way to simplify the
problem is to make the uniform-field approximation, where we replace $\Gi$ (i
= X, XX) in Eqs. (\ref{eq:ThreeLevel_N1})- (\ref{eq:P_abs}) with its spatially
averaged value
\begin{equation}\label{eq:Gamma_UFA}
\bar{\Gamma}_i = \frac{1}{V_m}\int d^3\rPOS \Gi = \frac{2g_o^2}{K_i}
\end{equation}
where $g_o = \mathbf{\mu}\cdot\mathbf{\hat{e}}\sqrt{\omega_c/2\hbar\epsilon_{o} \epsilon_{eff}V_m}$ and
\begin{equation}\label{eq:epsilon_eff}
\epsilon_{eff} = \frac{\int d^3\rPOS|E(\rPOS)|^2\epsilon(\rPOS)}{\int d^3\rPOS|E(\rPOS)|^2}
\end{equation}

Under the uniform field approximation the atomic population densities
$n_j(\rPOS)$ are no longer spatially varying.  We can therefore express the
equations of motion in terms of the total number of quantum dots in state $j$
given by $N_j = V_m n_j$ where $V_m$ is the cavity mode volume.  These
quantum dot populations must satisfy the constraint that $\sum_j N_j = N$,
where $N$ is the total number of quantum dots contained in the cavity.  With
these definitions, the equations of motion become the standard cavity-atom
rate equations, given by
\begin{eqnarray}
\label{eq:UniformApprox_N1}
{\pd {N}_1\over\pd t}&=&\barGX[(p + 1)({N}_2 + {N}_3) - 2p{N}_1] + \gamma_0 ({N}_2 + {N}_3) - 2R{N}_1\\
\label{eq:UniformApprox_N2}
{\pd {N}_2\over\pd t}&=&- \barGX[(p + 1){N}_2 - p{N}_1] + \barGXX[(p + 1){N}_4 - p{N}_2] - \gamma_0 {N}_2 + \gamma_2 {N}_4\nn\\ & + & R({N}_1 - {N}_2)\\
\label{eq:UniformApprox_N3}
{\pd {N}_3\over\pd t}&=&- \barGX[(p + 1){N}_3 - p{N}_1] + \barGXX[(p + 1){N}_4 - p{N}_3] - \gamma_0 {N}_3 + \gamma_2 {N}_4\nn\\ & + & R({N}_1 - {N}_3)\\
\label{eq:UniformApprox_N4}
{\pd {N}_4\over\pd t}&=&- \barGXX[2(p + 1){N}_4 - p({N}_2+{N}_3)] - 2\gamma_2 {N}_4 + R({N}_2+{N}_3)\\
\label{eq:UniformApprox_n}
{\pd p \over\pd t}&=&-p\kappa  + p\bar{G}(p) + \bar{\alpha}(p)
\end{eqnarray}
where
\begin{equation}\label{eq:G_UFA}
\bar{G}(p) = \barGX({N}_2 + {N}_3 -2{N}_1) + \barGXX(2{N}_4 - {N}_2 - {N}_3)
\end{equation}
and
\begin{equation}\label{eq:alpha_UFA}
\bar{\alpha}(p) = \barGX({N}_2 + {N}_3) + 2\barGXX{N}_4
\end{equation}
are the gain coefficient and spontaneous emission rate into the lasing mode. The absorbed power is given by
\begin{equation}\label{eq:P_abs_UFA}
\bar{P}_{abs} = \hbar\omega_{p}R(2{N}_1+ {N}_2+ {N}_3)
\end{equation}
The output power of the laser is still given by Eq.~(\ref{eq:P_out}).

We first determine the minimum number of quantum dots required to achieve
lasing.  We define $N_{th}$ as the total number of quantum dots in the cavity required to achieve a small signal gain equal to the cavity loss ($\lim_{p\to 0}\bar{G}(p) =
\kappa$), and calculate it by using the analytical steady-state
solutions to Eqs.~(\ref{eq:UniformApprox_N1})- (\ref{eq:UniformApprox_N4}) along with the condition $\sum_j N_j = N$ (see Appendices ~\ref{appendix:Expression_N_UFA},~\ref{appendix:Nth_UniformApprox}).
To perform calculations, we consider the specific example of colloidal
CdSe/ZnS core-shell quantum dots that emit in a wavelength range of
500-700 nm. We perform simulations using a dephasing rate of $\gamma_d =$
4.39$\times$10$^4$ ns$^{-1}$~\cite{VanSark2001}, a spontaneous emission rate
of $\gamma_0$ = $1/18$ ns$^{-1}$~\cite{Lounis2000}, and an Auger
recombination rate of $\gamma_a$ = $1/300$ ps$^{-1}$~\cite{Klimov2000a,Bruchez1998}. Nanocrystal quantum dots can be incorporated into photonic devices in a variety of ways such as spin-casting \cite{Min2006,Bose2007,Fushman2005,Rakher2011,Rakher2010} and immersion in liquid suspension ~\cite{Wu2007,Snee2005}. In these cases, the quantum dots reside on the surfaces of the devices, so we set $\epsilon_{eff} = 1$.

\begin{figure}[htbp]
\centering
\includegraphics[width=\textwidth]{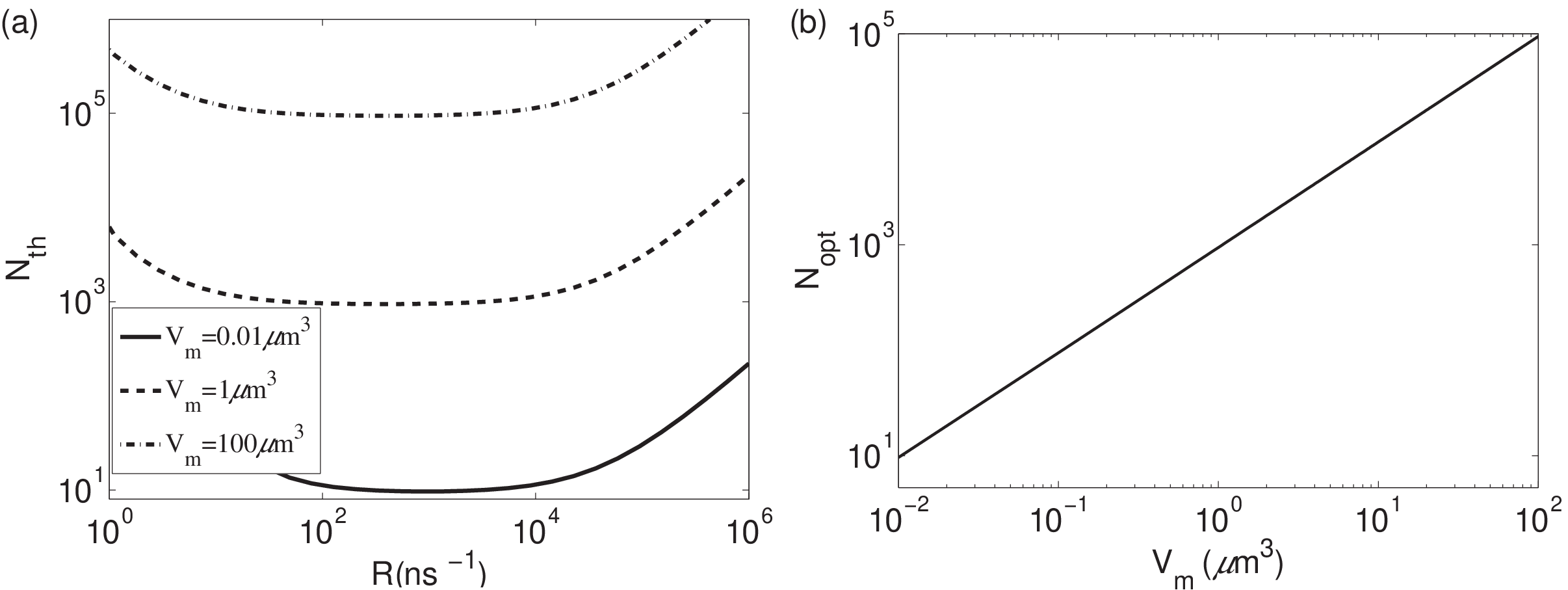}
\caption{(a) $N_{th}$ as a function of pump rate for $V_m = 0.01 \mu m^3$, 1 $\mu m^3$ and 100$ \mu m^3$, $\gamma_a$ = $1/300$ ps$^{-1}$. (b) $N_{opt}$ for different mode-volumes for $\gamma_a$ = $1/300$ ps$^{-1}$.}
\label{fig:Fig2_Nth_Nopt}
\end{figure}

Figure \ref{fig:Fig2_Nth_Nopt}(a) plots $N_{th}$ as a function of pump rate $R$ for $V_m = 0.01 \mu m^3$, 1$ \mu m^3$ and 100$ \mu m^3$ and $\gamma_a$ = $1/300$ ps$^{-1}$. Each mode-volume exhibits an optimum pump rate where the threshold quantum dot number is
minimum. We denote this minimum threshold quantum dot number by $N_{opt}$.
Figure \ref{fig:Fig2_Nth_Nopt}(b) plots $N_{opt}$ as a function of $V_m$. The
figure shows that $N_{opt}$ scales linearly with mode-volume.

\begin{figure}[htbp]
\centering
\includegraphics[width=\textwidth]{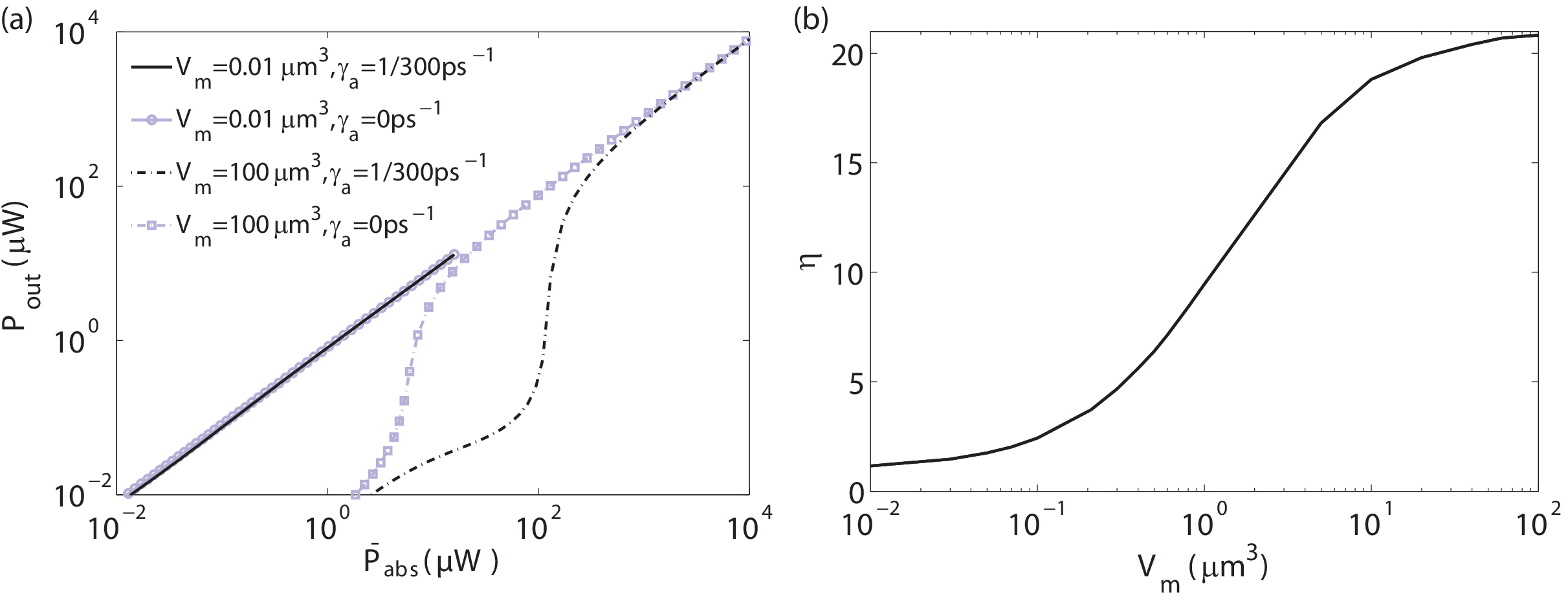}
\caption{(a) Laser output power as a function of the absorbed pump power for $V_m = 0.01 \mu m^3$ and 100$ \mu m^3$. (b)$\eta$ as a function of mode-volume.}
\label{fig:Fig3_UFA_LL_eta}
\end{figure}

Next, we investigate the laser input-output power characteristics. We
calculate the laser output power (using Eq. (\ref{eq:P_out})) and the absorbed
pump power (using Eq. (\ref{eq:P_abs_UFA})) using the numerical steady-state
solutions to Eqs. (\ref{eq:UniformApprox_N1})-(\ref{eq:UniformApprox_n}).
Figure~\ref{fig:Fig3_UFA_LL_eta}(a) plots $P_{out}$ as a function of
$\bar{P}_{abs}$ (also known as the light-in light-out curve), under the
uniform-field approximation, for two different mode-volumes of $V_m =$ 0.01
$\mu m^3$ and 100 $\mu m^3$, as well as two different Auger recombination
rates of $\gamma_a$ = $1/300$ ps$^{-1}$ and $0$. We set $Q =$ 20000 and $N =
2N_{opt}$ (Fig. ~\ref{fig:Fig2_Nth_Nopt}(b)) for each respective mode-volume.
We calculate the curves in Fig.~\ref{fig:Fig3_UFA_LL_eta}(a) using the same
range of $R$ values for both the mode-volumes.  We note that the curves for
the small mode volume cavity terminate earlier than those of the large mode
volume cavity because the number of quantum dots contained inside the cavity
mode-volume is much lower, which reduces the maximum output power.

The cavities with $V_m =$ 100$\mu$m$^{3}$, indicated by the dashed curves in
Fig.~\ref{fig:Fig3_UFA_LL_eta}(a), exhibit a pronounced threshold.
Near threshold, the light-in light-out curve takes on the well-known S-curve
behavior as it transitions from the below-threshold to above-threshold
regime. Auger recombination increases the threshold by quenching the gain,
which causes the S-curve region to occur at higher absorbed powers. Similar
to $N_{th}$, we define the threshold power as the absorbed power where the
small signal gain equals the cavity loss. We calculate this value numerically
using the steady state solutions to Eqs.~(\ref{eq:UniformApprox_N1})- (\ref{eq:UniformApprox_N4}), along with Eq. (\ref{eq:P_abs_UFA}). The threshold power for $V_m =$ 100 $\mu m^3$ is 122.7 $\mu$W when $\gamma_a$ = $1/300$ ps$^{-1}$, and
5.9 $\mu$W when $\gamma_a$ = 0. Auger recombination therefore increases the
lasing threshold by a factor of 21.  When the mode volume is $V_m =$ 0.01
$\mu m^3$ the light-in light-out curve exhibits a thresholdless lasing
behavior. The output power is nearly a linear function of the input power.
Using the same definition of threshold, we determine the threshold powers
with and without Auger recombination to be 97 nW and 84 nW respectively,
corresponding to an increase of only 1.2.  Thus, not only does the small mode
volume cavity exhibit a much lower overall lasing threshold, but the lasing
threshold is also largely unaffected by Auger recombination.

Figure \ref{fig:Fig3_UFA_LL_eta}(b) plots $\eta = P/P'$ as a
function of $V_m$, where $P$ is the absorbed pump power at threshold with
$\gamma_a$ = $1/300$ ps$^{-1}$ and $P'$ is the absorbed pump power at threshold with $\gamma_a$ = 0. We set the total quantum dot number in the cavities to $N = 2N_{opt}$
for each value of $V_m$ (Fig. ~\ref{fig:Fig2_Nth_Nopt}(b)). From this curve, we
observe that below a mode-volume of 0.1 $\mu m^3$ the lasing threshold is
largely unaffected by Auger recombination. Above this mode volume, $\eta$
rapidly increases and eventually reaches a saturated value. At large
mode-volumes, $\eta$ becomes independent of the mode volume itself and
achieves an asymptotic limit. From the upper and the lower limits of $\eta$ (21 and 1.2, respectively), we determine that spontaneous emission enhancement can reduce the lasing threshold up to a factor of 17.

\begin{figure}[htbp]
\centering
{\includegraphics[width=7cm]{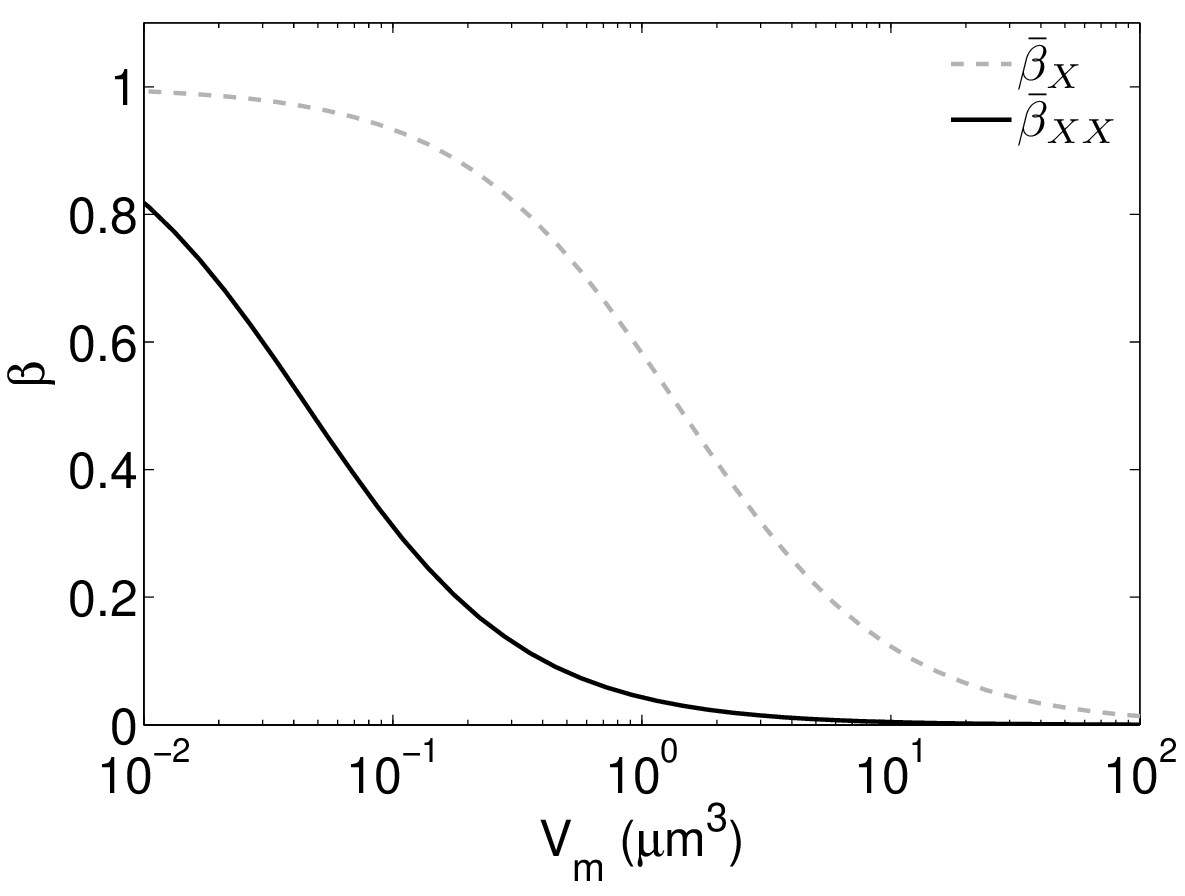}}
\caption{Spontaneous emission coupling efficiency for single-exciton transition $\bar\beta_{X}$ and biexciton transition $\bar\beta_{XX}$ as a function of $V_m$ for $\gamma_a$ = $1/300$ ps$^{-1}$}
\label{fig:Fig4_BetaVsVm_UFA}
\end{figure}

To verify that the improvement in lasing threshold is due to spontaneous emission
enhancement, we calculate the spontaneous emission coupling efficiency for
the exciton and biexciton transition as a function of $V_m$. Using the uniform field
approximation, we replace $\Gi$ (i = X, XX) in Eqs.
(\ref{eq:Beta_X}) - (\ref{eq:Beta_XX}) with its spatially averaged value
$\bar{\Gamma}_i$ which removes the spatial dependence and results in the
simplified expressions for the coupling efficiencies given by
\begin{eqnarray}
\label{eq:Beta_X_UFA}
\bar\beta_X &=& \frac{\barGX}{\barGX + \gamma_0}\\
\label{eq:Beta_XX_UFA}
\bar\beta_{XX} &=&\frac{\barGXX}{\barGXX + \gamma_2}
\end{eqnarray}

Figure~\ref{fig:Fig4_BetaVsVm_UFA} plots spontaneous emission coupling efficiencies
for the single-exciton transition $\bar\beta_X$ and the biexciton transition
$\bar\beta_{XX}$ as a function of $V_m$ using $\gamma_a$ = $1/300$ ps$^{-1}$.  At $V_m =$ 100 $\mu m^3$, $\bar\beta_{XX}$ is more than an order of magnitude smaller than $\bar\beta_{X}$.  As the mode
volume decreases the two efficiencies approach unity.  The coupling efficiency of the biexciton transition begins to increase sharply and approach unity around the same mode-volume where $\eta$ (Fig. \ref{fig:Fig3_UFA_LL_eta}(b)) begins to saturate to unity. Thus, at small mode-volumes $\bar\beta_{XX}$ is insensitive to Auger recombination, and therefore the threshold pump power does
not significantly change as indicated in Fig. \ref{fig:Fig3_UFA_LL_eta}(b).

\section{Cavity device structure for low-threshold laser}
\label{section:NanoBeamCavity}

The previous section established the advantage of using small mode-volume
cavities to achieve low threshold lasing with nanocrystal quantum dots. A
promising device structure for attaining this requirement is the nanobeam
photonic crystal cavity. Nanobeam photonic crystal cavities have been
previously studied in a variety of material systems, such as silicon
\cite{Foresi1997,Deotare2009,Quan2011}, silicon nitride
\cite{Eichenfield2009,Khan2011}, silicon dioxide
\cite{Velha2007,Zain2008,Gong2010}, and gallium arsenide
\cite{Rundquist2011,Rivoire2011}, and have been theoretically predicted to
achieve mode-volumes approaching the diffraction limit
~\cite{Deotare2009,Quan2011,Gong2010,Chan2009}.

\begin{figure}[htbp]
\centering
{\includegraphics[width=7cm]{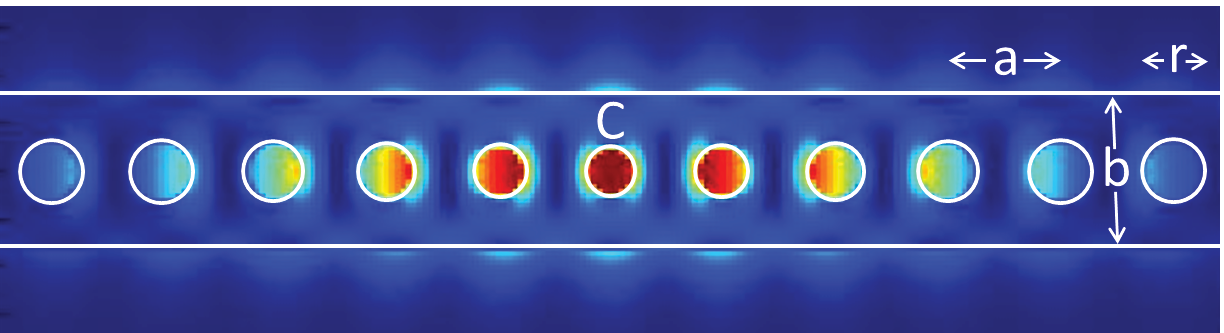}}
\caption{The electric field intensity ($|E|^2$) of the resonant cavity mode of a nanobeam photonic crystal cavity. The seven holes in the center form the cavity defect.}
\label{fig:Fig5_AirMode}
\end{figure}

Figure ~\ref{fig:Fig5_AirMode} shows the nanobeam photonic crystal cavity design
that we consider for low threshold lasing. Nanocrystal quantum dots are
typically spin cast onto the device and therefore reside outside the
dielectric. We therefore design the cavity mode to be localized in the air
holes rather than the dielectric material. This design choice maximizes the
field overlap with the quantum dots.

The structure is composed of a silicon nitride beam with a one-dimensional
periodic array of air holes (radius $r = 0.24a$, where $a$ is the lattice
constant). The cavity is composed of a defect in the structure created by
gradually reducing the radius of the three holes on either side of the hole
labelled C to a minimum of $r_0$ = 0.2$a$. The adiabatic reduction of hole
radius creates a smooth confinement for the photon and minimizes scattering
due to edge states~\cite{Quan2010}. The cavity is designed with beam
thickness $d$ = 0.727$a$ and beam width $b$ = 1.163$a$. The index of
refraction of silicon nitride is set to 2.01 \cite{Barth2007}. We calculate
the mode of the cavity using three dimensional finite-difference time-domain
simulation (Lumerical Solutions, Inc.). Figure~\ref{fig:Fig5_AirMode} shows the
calculated electric field intensity overlaid on the structure. The computed
mode-volume is $V_m$ = 0.38$\lambda^3$ (= 0.11 $\mu m^3$) and the quality factor is $Q = 64,000$.

Nanobeam photonic crystal cavities achieve mode-volumes that are on the order
of a cubic wavelength. When the confinement volume of the cavity approaches
the spatial variation of the field distribution, the uniform-field
approximation can break down. We therefore analyze the nanobeam laser both
with and without this approximation. We calculate $\epsilon_{eff} = 1.9$
for the cavity by numerically integrating Eq. (\ref{eq:epsilon_eff}) using
the computed electric field intensity profile of the simulated cavity
structure (Fig. \ref{fig:Fig5_AirMode}). Calculations under the uniform-field
approximation follow the same approach as in the section \ref{section:UFA}.

In order to investigate the input-output characteristics of the nanobeam
laser without the uniform-field approximation, we first determine the total
number of quantum dots required for achieving lasing threshold. We assume a
uniform volume-density of quantum dots, denoted by $n = N/V_p$ where $V_p$ is
the volume of the optically pumped region.  We assume quantum dots reside
only in the air holes and on the top of the nanobeam, which are optically
pumped with an illumination spot with a diameter of 690 nm, covering the central three holes of the cavity (Fig. \ref{fig:Fig5_AirMode}). We divide the illuminated volume into small volume elements (with volume $\Delta V$ at location $\rPOS$) and numerically
solve Eqs. (\ref{eq:ThreeLevel_N1})- (\ref{eq:ThreeLevel_N4}) and Eq.~(\ref{eq:G}) in steady state, along with the conditions $\sum_j n_j(\rPOS) = n$ for each volume element, and numerically determine the required $n$ to achieve $\lim_{p\to 0} G(p) = \kappa$. We assume that absorption loss due to quantum dots outside of the excitation volume are negligible compared to other loss mechanisms in the cavity.

Using the same simulation parameters as in the previous section, we
numerically calculate the minimum number of quantum dots required to achieve
threshold to be $N_{opt}= 60$.  This number is nearly identical to the value
calculated using the uniform-field approximation which is 62. Next, we
calculate the light-in light out curve using Eq. (\ref{eq:P_abs}) and Eq.
(\ref{eq:P_out}) without the uniform-field approximation.  As in the previous
section, we set the total number of quantum dots to be $N = 2N_{opt}$.

\begin{figure}[htbp]
\centering
\includegraphics[width=\textwidth]{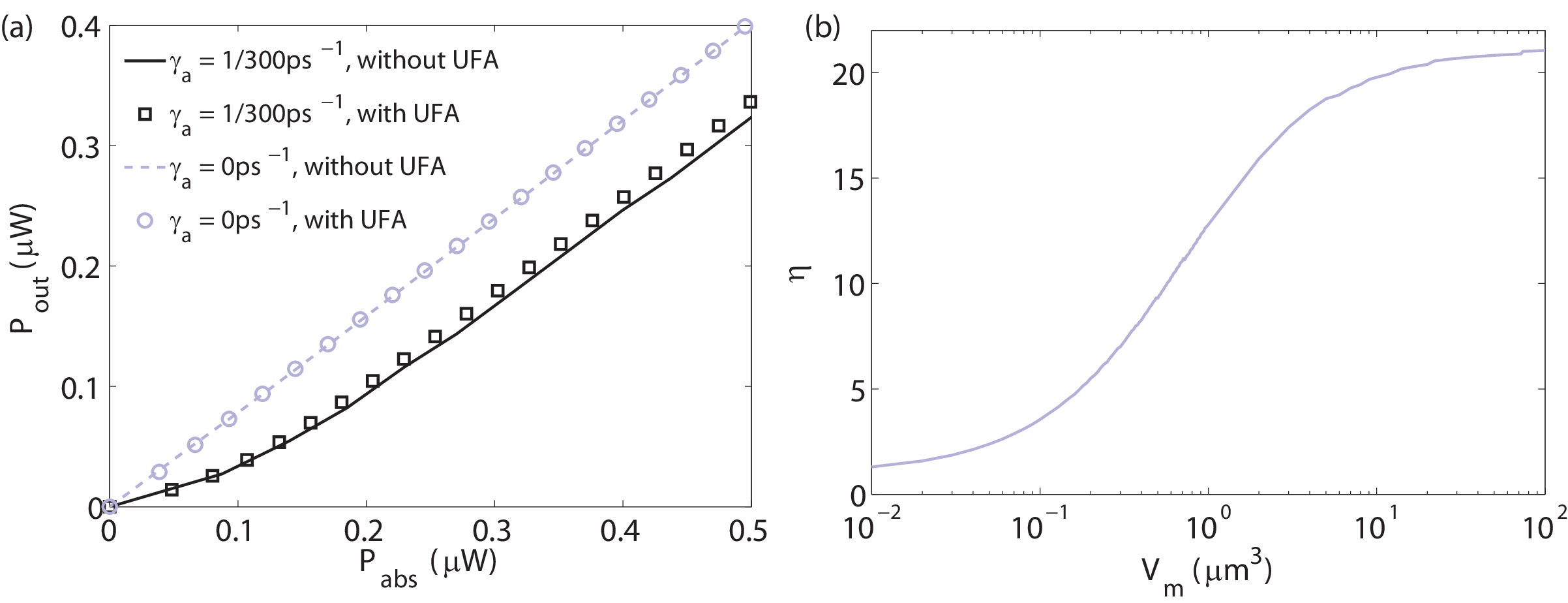}
\caption{(a) Output power as a function of the absorbed pump power for nanocrystal quantum dot laser comprised of nanobeam photonic crystal cavity, using $\gamma_a =$ $1/300$ ps$^{-1}$ and 0, both with and without uniform-field approximation (abbreviated as UFA in the legend). (b) $\eta$ as a function of mode-volume under the uniform-field approximation for $\epsilon_{eff} =$ 1.9 and $Q =$ 64,000.}
\label{fig:Fig6_ABNB_LL_eta}
\end{figure}

Figure \ref{fig:Fig6_ABNB_LL_eta}(a) plots $P_{out}$ as a function of
$P_{abs}$ for the nanobeam photonic crystal cavity with simulated $Q =
64,000$ using $\gamma_a =$ $1/300$ ps$^{-1}$ and
0, both with and without the uniform-field approximation.  The calculations
show good agreement between the predicted input-output characteristics of the
laser with and without the uniform-field approximation. Without the
uniform-field approximation, the absorbed pump power at threshold for the
nanobeam laser is 109.8 nW for $\gamma_a =$ $1/300$ ps$^{-1}$ and 29.9 nW for $\gamma_a =$ 0, resulting in $\eta$ = 3.7. With the uniform-field approximation, the absorbed pump power at threshold for the nanobeam laser is 112.6 nW for $\gamma_a
=$ $1/300$ ps$^{-1}$ and 30 nW for $\gamma_a =$ 0, resulting in $\eta$ = 3.8.

The $\epsilon_{eff}$ for the nanobeam cavity, calculated from the cavity-field distribution, is 1.9. This calculated $\epsilon_{eff}$ is higher than the unity assumption in the previous section because in this realistic cavity design a fraction of the cavity field leaks into the dielectric medium (Fig. \ref{fig:Fig5_AirMode}). Figure ~\ref{fig:Fig6_ABNB_LL_eta}(b) plots $\eta$ as a function of $V_m$ under the uniform-field approximation for the same parameters used in Fig. ~\ref{fig:Fig6_ABNB_LL_eta}(a). For a cavity with a mode volume of 100 $\mu m^3$, we determine that $\eta =$ 21.1.  This value is 5.6 times larger than the value for the nanobeam cavity. Thus, the nanobeam cavity lasing threshold is much less sensitive to Auger recombination. 

\section{Conclusion}
In conclusion, we have theoretically shown that cavity-enhanced spontaneous
emission of the biexciton reduces the effect of Auger recombination, leading
to a lower lasing threshold. We developed a numerical model for a laser composed of an ensemble of nanocrystal quantum dots coupled to an optical cavity. The model can be expanded to incorporate more complex behavior of nanocrystal quantum dots, such as blinking, by introducing additional trap states into the quantum dot level structure
~\cite{Zhao2010,Kuno2003}. This model can also be used to study lasing with
other room-temperature emitters such as quantum rods ~\cite{Kazes2002,HtoonPRL2003}, and other types of cavities such as plasmonic apertures ~\cite{Choy2011}. Our results provide a direction for development of low-threshold and highly tunable nanolasers that use nanocrystal quantum dot as gain material at room temperature.

\bigskip
\appendix{\bf{Appendices}}

\section{Liouvillian superoperator L}
\label{appendix:Liouvillian}
The Liouvillian superoperator $\mathbf{L}$ can be expressed as $\mathbf{L} = \mathbf{L_{NQD}}+\mathbf{L_{pump}}+\mathbf{L_{cavity}}$, where $\mathbf{L_{NQD}}$ accounts for the spontaneous relaxation of the quantum dot level structure, $\mathbf{L_{pump}}$ accounts for the incoherent pumping of the quantum dot population, and $\mathbf{L_{cavity}}$ accounts for the cavity decay. These operators are
\begin{eqnarray}\label{eq:L_NQD}
\mathbf{L_{NQD}}\rho  = &\sum_{m=1}^{N}&\frac{\gamma_{0,m}}{2}(2\sigma_{12,m}\rho\sigma_{21,m} - \sigma_{21,m}\sigma_{12,m}\rho  - \rho\sigma_{21,m}\sigma_{12,m} \nn\\
&+& 2\sigma_{13,m}\rho\sigma_{31,m} -\sigma_{31,m}\sigma_{13,m}\rho - \rho\sigma_{31,m}\sigma_{13,m})\nn\\
 &+& \frac{\gamma_{2,m}}{2}(2\sigma_{24,m}\rho\sigma_{42,m} - \sigma_{42,m}\sigma_{24,m}\rho - \rho\sigma_{42,m}\sigma_{24,m}\nn\\
&+& 2\sigma_{34,m}\rho\sigma_{43,m} - \sigma_{43,m}\sigma_{34,m}\rho - \rho\sigma_{43,m}\sigma_{34,m})
\end{eqnarray}
\begin{eqnarray}\label{eq:L_pump}
\mathbf{L_{pump}}\rho  = &\sum_{m=1}^{N}&\frac{R}{2}(2\sigma_{21,m}\rho\sigma_{12,m} - \sigma_{12,m}\sigma_{21,m}\rho - \rho\sigma_{12,m}\sigma_{21,m}\nn\\
&+& 2\sigma_{31,m}\rho\sigma_{13,m} - \sigma_{13,m}\sigma_{31,m}\rho - \rho\sigma_{13,m}\sigma_{31,m}\nn\\
&+& 2\sigma_{42,m}\rho\sigma_{24,m} - \sigma_{24,m}\sigma_{42,m}\rho - \rho\sigma_{24,m}\sigma_{42,m}\nn\\
&+& 2\sigma_{43,m}\rho\sigma_{34,m} - \sigma_{34,m}\sigma_{43,m}\rho - \rho\sigma_{34,m}\sigma_{43,m})
\end{eqnarray}
\begin{eqnarray}\label{eq:L_cavity}
{\mathbf{L_{cavity}}\rho = \frac{\kappa}{2}(2\mathbf{a}\rho\mathbf{a}^\dagger - \mathbf{a}^\dagger\mathbf{a}\rho - \rho\mathbf{a}^\dagger\mathbf{a})}
\end{eqnarray}
The cavity energy decay rate is $\kappa = \omega_c/Q$.

\section{Equations of motion: projected on quantum dot levels}
\label{appendix:Equations_N}
The equations of motion for the projections of $\rho$ on  the levels (ij) of the $m^{th}$ quantum dot and photon states (pp') $\rho_{ip,jp'}^{m} = _m\langle i,p|\rho|j,p'\rangle_m$ (i,j = 1 , 2, 3, 4) and ($p,p'= 0$ to $\infty$)  are obtained using Eq. (\ref{eq:MasterEquation}):
\begin{eqnarray}\label{eq:PhotonEqn_Motion11}
\frac{\partial\rho_{1p,1p}^{m}}{\partial t} &=& ig_{m}\sqrt{p}(\rho_{1p,2p-1}^{m} - \rho_{2p-1,1p}^{m} + \rho_{1p,3p-1}^{m} - \rho_{3p-1,1p}^{m}) - 2R\rho_{1p,1p}^{m} \nn\\
&+& \gamma_{0}(\rho_{2p,2p}^{m} + \rho_{3p,3p}^{m} )+ \kappa((p+1)\rho_{1p+1,1p+1}^{m} - p\rho_{1p,1p}^{m})
\end{eqnarray}
\begin{eqnarray}\label{eq:PhotonEqn_Motion22}
\frac{\partial\rho_{2p,2p}^{m}}{\partial t} &=& ig_{m}(\sqrt{p+1}(\rho_{2p,1p+1}^{m} - \rho_{1p+1,2p}^{m}) + \sqrt{p}(\rho_{2p,4p-1}^{m} - \rho_{4p-1,2p}^{m})) \nn\\
&-& (\gamma_{0} +  R)\rho_{2p,2p}^{m} + R\rho_{1p,1p}^{m} + \gamma_{2}\rho_{4p,4p}^{m} + \kappa((p+1)\rho_{2p+1,2p+1}^{m} - p\rho_{2p,2p}^{m})
\end{eqnarray}
\begin{eqnarray}\label{eq:PhotonEqn_Motion33}
\frac{\partial\rho_{3p,3p}^{m}}{\partial t} &=& ig_{m}(\sqrt{p+1}(\rho_{3p,1p+1}^{m} - \rho_{1p+1,3p}^{m}) + \sqrt{p}(\rho_{3p,4p-1}^{m} - \rho_{4p-1,3p}^{m})) \nn\\
&-& (\gamma_{0} +  R)\rho_{3p,3p}^{m}+ R\rho_{1p,1p}^{m} + \gamma_{2}\rho_{4p,4p}^{m} + \kappa((p+1)\rho_{3p+1,3p+1}^{m} - p\rho_{3p,3p}^{m})
\end{eqnarray}
\begin{eqnarray}\label{eq:PhotonEqn_Motion44}
\frac{\partial\rho_{4p,4p}^{m}}{\partial t} &=& ig_{m}\sqrt{p+1}(\rho_{4p,2p+1}^{m} - \rho_{2p+1,4p}^{m} + \rho_{4p,3p+1}^{m} - \rho_{3p+1,4p}^{m}) - 2\gamma_{2}\rho_{3p,3p}^{m} \nn\\
&+& R(\rho_{2p,2p}^{m} + \rho_{3p,3p}^{m})+ \kappa((p+1)\rho_{3p+1,3p+1}^{m} - p\rho_{3p,3p}^{m})
\end{eqnarray}
\begin{eqnarray}\label{eq:PhotonEqn_Motion12}
\frac{\partial\rho_{1p,2p-1}^{m}}{\partial t} = ig_{m}\sqrt{p}(\rho_{1p,1p}^{m} - \rho_{2p-1,2p-1}^{m}) - K_{X}\rho_{1p,2p-1}^{m}
\end{eqnarray}
\begin{equation}\label{eq:PhotonEqn_Motion24}
\frac{\partial\rho_{2p,4p-1}^{m}}{\partial t} = ig_{m}\sqrt{p}(\rho_{2p,2p}^{m} - \rho_{4p-1,4p-1}^{m}) - K_{XX}\rho_{2p,4p-1}^{m}
\end{equation}
\begin{equation}\label{eq:PhotonEqn_Motion13}
\frac{\partial\rho_{1p,3p-1}^{m}}{\partial t} = ig_{m}\sqrt{p}(\rho_{1p,1p}^{m} - \rho_{3p-1,3p-1}^{m}) - K_{X}\rho_{1p,3p-1}^{m}
\end{equation}
\begin{equation}\label{eq:PhotonEqn_Motion34}
\frac{\partial\rho_{3p,4p-1}^{m}}{\partial t} = ig_{m}\sqrt{p}(\rho_{3p,3p}^{m} - \rho_{4p-1,4p-1}^{m}) - K_{XX}\rho_{3p,4p-1}^{m}
\end{equation}
Here, $K_{X}\;= (\gamma_{0} + \gamma_{d} + 3R)/2$ and $K_{XX}\;= (\gamma_{0} + 2\gamma_{2} + \gamma_{d} + R)/2$ are the total relaxation rates of the diagonal terms, and $\gamma_{d}$ is the dephasing rate of the quantum dot (added phenomenologically). We set dephasing rate to be much greater than the cavity decay rate $\gamma_d \gg \kappa$, allowing us to drop the cavity decay contributions from the equations of motion of off-diagonal terms (Eqs. (\ref{eq:PhotonEqn_Motion12}) - (\ref{eq:PhotonEqn_Motion34})). Large dephasing rate also allows us to adiabatically eliminate the expectation value $\langle\rho_{ip,jp'}\rangle$ of the off-diagonal terms ($i \not= j$)from Eqs. (\ref{eq:PhotonEqn_Motion12}) - (\ref{eq:PhotonEqn_Motion34}), and reduces Eqs. (\ref{eq:PhotonEqn_Motion11}) - (\ref{eq:PhotonEqn_Motion44}) to
\begin{eqnarray}\label{eq:FinalPhotonEqn_Motion11}
\frac{\partial\rho_{1p,1p}^{m}}{\partial t} &=& \frac{2g^{2}_{m}}{K_{X}}(\rho_{2p-1,2p-1}^{m} + \rho_{3p-1,3p-1}^{m}-2\rho_{1p,1p}^{m})p - 2R\rho_{1p,1p}^{m} \nn\\
&+& \gamma_{0}(\rho_{2p,2p}^{m}+\rho_{3p,3p}^{m})+ \kappa((p+1)\rho_{1p+1,1p+1}^{m} - p\rho_{1p,1p}^{m})
\end{eqnarray}
\begin{eqnarray}\label{eq:FinalPhotonEqn_Motion22}
\frac{\partial\rho_{2p,2p}^{m}}{\partial t} &=& -\frac{2g^{2}_{m}}{K_{X}}(\rho_{2p,2p}^{m} - \rho_{1p+1,1p+1}^{m})(p+1) +\frac{2g^{2}_{m}}{K_{XX}}( \rho_{4p-1,4p-1}^{m} - \rho_{2p,2p}^{m} )p \nn\\
&-& (\gamma_{0} +  R)\rho_{2p,2p}^{m}+\; R\rho_{1p,1p}^{m} + \gamma_{2}\rho_{4p,4p}^{m} + \kappa((p+1)\rho_{2p+1,2p+1}^{m} - p\rho_{2p,2p}^{m})
\end{eqnarray}
\begin{eqnarray}\label{eq:FinalPhotonEqn_Motion33}
\frac{\partial\rho_{3p,3p}^{m}}{\partial t} &=& -\frac{2g^{2}_{m}}{K_{X}}(\rho_{3p,3p}^{m} - \rho_{1p+1,1p+1}^{m})(p+1) +\frac{2g^{2}_{m}}{K_{XX}}( \rho_{4p-1,4p-1}^{m} - \rho_{3p,3p}^{m} )p \nn\\
&-& (\gamma_{0} +  R)\rho_{3p,3p}^{m}+\; R\rho_{1p,1p}^{m} + \gamma_{2}\rho_{4p,4p}^{m} + \kappa((p+1)\rho_{3p+1,3p+1}^{m} - p\rho_{3p,3p}^{m})
\end{eqnarray}
\begin{eqnarray}\label{eq:FinalPhotonEqn_Motion44}
\frac{\partial\rho_{4p,4p}^{m}}{\partial t} &=& -\frac{2g^{2}_{m}}{K_{XX}}(2\rho_{4p,4p}^{m} - \rho_{2p+1,2p+1}^{m}- \rho_{3p+1,3p+1}^{m})(p+1)- 2\gamma_{2}\rho_{4p,4p}^{m} \nn\\
&+& R(\rho_{2p,2p}^{m}+\rho_{3p,3p}^{m})+ \kappa((p+1)\rho_{4p+1,4p+1}^{m} - p\rho_{4p,4p}^{m})
\end{eqnarray}
Now, tracing over all the photon states in Eq. (\ref{eq:FinalPhotonEqn_Motion11}) - (\ref{eq:FinalPhotonEqn_Motion44}), and  applying semi-classical approximation to factorize full density matrix element into quantum dot and field parts such that $\rho_{ip,ip} = \rho_{ii}\rho_{pp}$, we get
\begin{equation}\label{eq:NQDEqn_Motion11}
\frac{\partial\rho_{11}^{m}}{\partial t} = \frac{2g^{2}_{m}}{K_{X}}(\rho_{22}^{m}+\rho_{33}^{m}-2\rho_{11}^{m})\langle p \rangle + \frac{2g^2}{K_{X}}(\rho_{22}^{m}+\rho_{33}^{m}) - 2R\rho_{11}^{m} + \gamma_{0}(\rho_{22}^{m}+\rho_{33}^{m})
\end{equation}
\begin{eqnarray}\label{eq:NQDEqn_Motion22}
\frac{\partial\rho_{22}^{m}}{\partial t} &=& -\frac{2g^{2}_{m}}{K_{X}}(\rho_{22}^{m}-\rho_{11}^{m})\langle p \rangle + \frac{2g^{2}_{m}}{K_{XX}}(\rho_{44}^{m}-\rho_{22}^{m})\langle p \rangle - \frac{2g^{2}_{m}}{K_{X}}\rho_{22}^{m} + \frac{2g^{2}_{m}}{K_{XX}}\rho_{44}^{m} \nn\\
&-& (\gamma_{0} +  R)\rho_{22}^{m} + R\rho_{11}^{m} + \gamma_{2}\rho_{44}^{m}
\end{eqnarray}
\begin{eqnarray}\label{eq:NQDEqn_Motion33}
\frac{\partial\rho_{33}^{m}}{\partial t} &=& -\frac{2g^{2}_{m}}{K_{X}}(\rho_{33}^{m}-\rho_{11}^{m})\langle p \rangle + \frac{2g^{2}_{m}}{K_{XX}}(\rho_{44}^{m}-\rho_{33}^{m})\langle p \rangle - \frac{2g^{2}_{m}}{K_{X}}\rho_{33}^{m} + \frac{2g^{2}_{m}}{K_{XX}}\rho_{44}^{m} \nn\\
&-& (\gamma_{0} +  R)\rho_{33}^{m} + R\rho_{11}^{m} + \gamma_{2}\rho_{44}^{m}
\end{eqnarray}
\begin{eqnarray}\label{eq:NQDEqn_Motion44}
\frac{\partial\rho_{44}^{m}}{\partial t} &=& -\frac{2g^{2}_{m}}{K_{XX}}(2\rho_{44}^{m}-\rho_{22}^{m}-\rho_{33}^{m})\langle p \rangle -\frac{4g^{2}_{m}}{K_{XX}}\rho_{44}^{m} - 2\gamma_{2}\rho_{44}^{m} + R(\rho_{22}^{m}+\rho_{33}^{m})
\end{eqnarray}
where $\langle p \rangle = \sum_{p}p\rho_{pp}$ is the mean photon number. We define $n_j(\rPOS) = \lim_{\Delta V\to 0}\sum_m\langle\sigma_{jj}^m\rangle/\Delta V$ as the quantum dot population density of the j$^{th}$ lasing level where the sum is carried out over all quantum dots contained in small volume $\Delta V$ at location $\rPOS$ and get Eqs. (\ref{eq:ThreeLevel_N1}) - (\ref{eq:ThreeLevel_N4}).

\section{Rate equation for mean cavity photon number}
\label{appendix:Equations_n}
The rate equation for the mean cavity photon number is given by
\begin{equation}\label{eq:PhotonRateDefn}
{\langle\dot{p}\rangle} = \sum_{p}p{\dot{\rho}_{pp}}
\end{equation}
Using Eqs. (\ref{eq:FinalPhotonEqn_Motion11}) - (\ref{eq:FinalPhotonEqn_Motion44})
\begin{eqnarray}\label{eq:PhotonRateDerivation}
{\langle\dot{p}\rangle} = \sum_{m}\sum_{p}p&\lbrace& \frac{2g^2_m}{K_{X}}(\rho_{2p-1,2p-1}^{m}-\rho_{1p,1p}^{m})p - \frac{2g^2_{m}}{K_{X}}(\rho_{2p,2p}^{m} - \rho_{1p+1,1p+1}^{m})(p+1)\nn\\
&+&\frac{2g^2_m}{K_{X}}(\rho_{3p-1,3p-1}^{m}-\rho_{1p,1p}^{m})p - \frac{2g^2_{m}}{K_{X}}(\rho_{3p,3p}^{m} - \rho_{1p+1,1p+1}^{m})(p+1)\nn\\
&+&\frac{2g^2_{m}}{K_{XX}}(\rho_{4p-1,4p-1}^{m} - \rho_{2p,2p}^{m})p - \frac{2g^2_{m}}{K_{XX}}(\rho_{4p,4p}^{m} - \rho_{2p+1,2p+1}^{m})(p+1)\nn\\
&+&\frac{2g^2_{m}}{K_{XX}}(\rho_{4p-1,4p-1}^{m} - \rho_{3p,3p}^{m})p - \frac{2g^2_{m}}{K_{XX}}(\rho_{4p,4p}^{m} - \rho_{3p+1,3p+1}^{m})(p+1)\nn\\
&-&\kappa(p\rho_{pp} - (p+1)\rho_{p+1p+1})\rbrace
\end{eqnarray}
Applying semi-classical approximation to factorize full density matrix element into quantum dot and field parts $\rho_{ip,ip} = \rho_{ii}\rho_{pp}$, and identifying $\sum_{p=0}^{\infty}p\rho_{p,p} = \langle p\rangle$ gives
\begin{eqnarray}\label{eq:Final_n_Rate}
{\langle\dot{p}\rangle} &=& - \kappa\langle p\rangle + \sum_{m}\lbrace\frac{2g^2_m}{K_{X}}(\rho_{22}^{m}+\rho_{33}^{m} - 2\rho_{11}^{m})\langle p\rangle + \frac{2g^2_m}{K_{XX}}(2\rho_{44}^{m} - \rho_{22}^{m}-\rho_{33}^{m})\langle p\rangle \nn\\
&+& \frac{2g^2_m}{K_{X}}(\rho_{22}^{m}+\rho_{33}^{m}) + \frac{4g^2_m}{K_{XX}}\rho_{44}^{m}\rbrace
\end{eqnarray}
Eq. (\ref{eq:Final_n_Rate}) leads us to Eq. (\ref{eq:ThreeLevel_n}).

\section{Expression for $N_j$ under the uniform-field approximation}
\label{appendix:Expression_N_UFA}
Assuming total number of quantum dots in the cavity, $N$, such that $\sum_i N_i = N$, Eqs. ~(\ref{eq:UniformApprox_N1})-(\ref{eq:UniformApprox_N4}) can be solved in the steady-state as
\begin{eqnarray}
\label{eq:UniformApproxExp_N1}N_{1} & = & \left(\frac{(p+1)\barGX + \gamma_0}{p\barGX + R}\right)\frac{N}{2\zeta}\\
\label{eq:UniformApproxExp_N23}N_{2} & = & N_{3} =  \frac{N}{2\zeta}\\
\label{eq:UniformApproxExp_N4}N_{4} & = & \left(\frac{p\barGXX + R}{(p+1)\barGXX + \gamma_2}\right)\frac{N}{2\zeta}\\
\label{eq:UniformApproxExp_zeta}\zeta & = & \frac{(p+1)\barGX + \gamma_0}{2(p\barGX + R)} + 1 + \frac{p\barGXX +
R}{2((p+1)\barGXX + \gamma_2)}
\end{eqnarray}
where $\zeta$ is the ratio of the total quantum dot population to the total single-exciton quantum dot population.

\section{Quantum dot number required for achieving lasing threshold}
Under uniform-field approximation
\label{appendix:Nth_UniformApprox}
\begin{equation}\label{eq:Nth_UniformApprox}
    N_{th} = \frac{\omega_c}{Q}\left(\frac{\frac{\barGX + \gamma_0}{2R} + 1 + \frac{R}{2\barGXX + 2\gamma_2}}{\barGX(1 - \frac{\barGX + \gamma_0}{R}) + \barGXX(\frac{R}{\barGXX + \gamma_2} - 1)}\right)
  \end{equation}
  
\section*{Acknowledgment}
We acknowledge funding support from the Physics Frontier Center at the Joint Quantum Institute (grant number PHY-0822671).

%%%%%%%%%%%%%%%%%%%%%%%%%%%%%%%%%%

%%%%%%%%%%%%%%%%%%%%%%%%%%%%%%%%%%%
%\bibliography{References}
\end{document}